\documentclass[aps,prl,twocolumn,showpacs,superscriptaddress,groupedaddress,preprintnumbers]{revtex4}  
\usepackage{graphicx}  
\usepackage{dcolumn}   
\usepackage{bm}        
\usepackage{amssymb}   
\usepackage{amsmath}
\usepackage{color}
\usepackage{lipsum}
\newcommand{\mpl}{M_{\text{Pl}}}

\newcommand{\be}{\begin{equation}}
\newcommand{\ee}{\end{equation}}
\newcommand{\bea}{\begin{eqnarray}}
\newcommand{\eea}{\end{eqnarray}}
\newcommand{\nn}{\nonumber}

\begin{document}

\preprint{IPMU18-0150}
\title{Quintessence Saves Higgs Instability}
\author{Chengcheng Han${}^a$, Shi Pi${}^a$ and Misao Sasaki${}^{a,b,c}$\\
\it
$^{a}$ Kavli Institute for the Physics and Mathematics of the Universe (WPI),
Chiba 277-8583, Japan\\
$^{b}$ Yukawa Institute for Theoretical Physics, Kyoto University,
Kyoto 606-8502, Japan\\
$^{c}$ Leung Center for Cosmology and Particle Astrophysics, National Taiwan University,
Taipei 10617, Taiwan
}
\date{\today}

\begin{abstract}
We study a model where quintessence potential $e^{-\xi\phi}$ coupled to Higgs potential. We calculate the evolution of the quintessence, and track the running of the effective Higgs self coupling. We find it slightly larger than that of the standard model in the past. Requiring the electroweak vacuum to be absolutely stable in inflationary era, we find a lower bound $\xi> 0.35\pm 0.05$, where the uncertainty is mainly from the measurement of the top quark mass. This lower bound, together with the upper bound from the observation for dark energy $\xi\lesssim0.6$, narrows down the parameter space and makes it possible to test this model in the near future. Interestingly, the bound on $\xi$, if actually shown to be the case by observation, supports the recently proposed Swampland Conjecture. 
\end{abstract}

\maketitle


\textit{Introduction}~
The accelerated expansion of our universe may be the most important discovery in cosmology in the last 30 years~\cite{Riess:1998cb,Perlmutter:1998np}. 
This requires that the energy density of the universe is dominated by some kind of dark energy with equation of motion $w_{\text{DE}}$ around $-1$. 
Currently whether the dark energy is a cosmological constant or a slow-rolling scalar field dubbed ``quintessence'' is still not clear, and more accurate 
observational data are expected. A small cosmological constant may be realized in Type IIB string theory by KKLT 
construction~\cite{Kachru:2003aw,Kachru:2003sx}. A quintessence field is a scalar field $\phi$ minimally coupled to gravity with a relatively flat potential, 
and the slow-rolling $\phi$ can drive the universe to accelerate\cite{Copeland:1997et,Zlatev:1998tr,Ratra:1987rm}. A typical quintessence potential 
of $e^{-\xi\phi}$ has been studied earlier in \cite{Ratra:1987rm,Wetterich:1987fm,Wetterich:1994bg,Lucchin:1984yf,Halliwell:1986ja,Ferreira:1997au,Ferreira:1997hj,Copeland:1997et,Caldwell:1997ii,Tsujikawa:2013fta}, 
and analyzed recently to constrain the parameter $\xi$ by the recent observational data~\cite{Agrawal:2018own,Heisenberg:2018yae,Akrami:2018ylq,Heisenberg:2018rdu,Cicoli:2018kdo,Marsh:2018kub,Murayama:2018lie}. 

There is also an accelerated expansion of the universe in the past, the primordial inflation~~\cite{Brout:1977ix,Sato:1980yn,Guth:1980zm,Linde:1981mu,Albrecht:1982wi}. During inflation, the Hubble parameter $H_\text{inf}$ is 
nearly a constant of order $\sim10^{13}$GeV, and any scalar field whose mass is smaller than $H_\text{inf}$ will have quantum fluctuations 
of order $H_\text{inf}$. The rapidly expanding universe can bring the quantum fluctuations generated deep inside the Hubble horizon to 
cosmological scales, which seeds as the initial condition for the primordial scalar and tensor perturbations we need to explain the cosmic 
microwave background anisotropy and the large scale structure today~\cite{Starobinsky:1979ty,Mukhanov:1981xt}.

The only scalar field we have observed in Nature is the Higgs boson~\cite{Englert:1964et,Higgs:1964pj}. The self coupling of Higgs boson receives quantum
 corrections from itself and the other standard model (SM) particles~\cite{Sher:1988mj,Degrassi:2012ry,EliasMiro:2011aa}. According to the best-fit 
 observational data of the Higgs boson mass \cite{Aad:2015zhl} and the top quark mass \cite{ATLAS:2014wva}, the Higgs self coupling would become 
 negative at an energy scale around $\Lambda_{\text{ins}}\sim10^{11}\text{GeV}$, with absolute stable electroweak (EW) vacuum excluded at $99\%$ 
 confidence level (CL). This may raise a serious problem during inflation since the quantum fluctuations of the Higgs boson becomes 
 very large to render the probability of ending up in the true vacuum substantially high, which means our universe is a real miracle.  

Here in this paper, we study a recently proposed model of quintessence $\phi$ which coupled to the Higgs boson ($\mpl=1$)~\cite{Denef:2018etk} 
\be\label{potential}
V(\phi,\mathcal{H})=e^{-\xi\phi}\left(\lambda\left(| \mathcal{H}|^2-v^2\right)^2+\Lambda\right)\;\;(\xi>0),
\ee
where  $\mathcal H =  \left( {\begin{array}{c}   0  \\  h/\sqrt2+ v  \\  \end{array} } \right) $, $v$ is the Higgs vacuum expectation value (VEV), 
$\xi$ is a coupling constant, $\Lambda$ is the current effective cosmological constant, and we set the value of the present quintessence field $\phi_0$ to 
be 0 for convenience. It can pass the test of equivalence principle according to the discussions in~\cite{Denef:2018etk}, and it is difficult to be 
detected in collider experiments since the interactions are suppressed by the Planck mass. 
Here in this paper our main concern is the Higgs instability during inflation. We compute the correction to the Higgs self coupling $\lambda$ 
from the evolution of the dilaton-like coefficient back in time $e^{-\xi\phi}$ up to the inflation era, and consider the condition on $\xi$ for the Higgs stability. 
Requiring that the Higgs potential to be absolutely stable during inflation, we obtain a lower bound of $\xi$. We find it it detectable in the near future. 

\textit{Higgs instability}~
The running of the Higgs self coupling is determined by the renormalization group equation (RGE),
\begin{eqnarray}
\frac{d \lambda}{d \ln \mu^2}&=&\frac{1}{(4\pi)^2} \left[   \lambda\left(12 \lambda+ 6 y_t^2 -\frac{9}{2}g_2^2 -\frac{3}{2}g_1^2\right)\right.  \nonumber \\
&&\left.- 3 y_t^4+\frac{9g_2^4}{16} +\frac{3 g_1^4}{16}+ \frac{3 g_2^2 g_1^2}{8} \right],
\end{eqnarray}
where $y_t$ is the top quark Yukawa coupling, $g_1$ and $g_2$ are the gauge couplings of $\text{SU}(2)_L$ and $\text{U}(1)_Y$, respectively.
The top Yukawa coupling $-3 y_t^4$ would dominate the running, which decreases with the energy scale $\mu$. 
The current best-fit data for the Higgs boson mass \cite{Aad:2015zhl} and the top quark mass \cite{ATLAS:2014wva} indicates that the self coupling 
becomes negative above an instability scale at around $\Lambda_\text{ins}\sim10^{11}$GeV. 
Since the lifetime of the metastable EW vacuum is much longer than the age of our universe, we do not have to worry about the doomsday of our current universe.
However, during inflation, the quantum fluctuation of Higgs boson is of order $H_\text{inf}$, and our patch of the universe will be kicked out 
to the true vacuum with negative energy density just in one $e$-fold. There are many papers discussing this issue~\cite{Espinosa:2007qp,EliasMiro:2012ay,Lebedev:2012sy,Fairbairn:2014zia,Branchina:2014usa,Bezrukov:2014ipa,Herranen:2014cua,Hook:2014uia,Kamada:2014ufa,Espinosa:2015qea,Ballesteros:2015iua,George:2015nza,Gong:2015gxf,Kawasaki:2016ijp,Ema:2016ehh,East:2016anr,Saha:2016ozn,Espinosa:2017sgp,Espinosa:2018eve}, including introducing some heavy scalar, or requiring a high reheating temperature, to save our universe from the instability. 
Here we try to solve this problem in the framework of the quintessence-Higgs coupling (\ref{potential}).

In this model, the quintessence couples to the Higgs sector and effectively changes its self coupling as $\lambda^\prime = \lambda e^{-\xi \phi}$, 
while leaving the other SM interactions intact. At present the Higgs sector is the same as the standard model. However, in the early universe, the 
quintessence $\phi$ would have had a larger negative value, which would enhance the value of the Higgs self coupling. 
To discuss the Higgs stability at high energy scales, when $h \gg v$, we define the effective potential as
\begin{eqnarray}
V_\text{eff}(h)=\lambda_\text{eff}(h) \frac{h^4}{4},
\end{eqnarray}
where the renormalization scale $\mu$ is set to be $h$. $\lambda_\text{eff}(h)$ can be written as
\begin{eqnarray}
\lambda_\text{eff}(h) = e^{4\Gamma (h)} \left[  \lambda^\prime( h) + \lambda^{(1)}(  h) + \lambda^{(2)}( h)   \right] 
\end{eqnarray}
Where $\lambda^\prime(h)$ is the value of $\lambda^\prime$ at the renormalization scale $ \mu=h$. 
It is derived from the RGE running of $\lambda^\prime$ with the initial value $\lambda^\prime(v) \equiv \lambda(v) e^{-\xi \phi}$. 
And $\Gamma (h) \equiv \int^h_{M_t} \gamma(\bar \mu) d \log \bar\mu$ where $\gamma$ is the anomalous dimension of the Higgs field. 
$\lambda^{(1)}(h)$, $\lambda^{(2)}(h)$ are the one-loop and two-loop corrections of  $\lambda_\text{eff}(h)$, respectively. 
In our calculation, we only count the effects from $y_t$, $g_1$, $g_2$, $g_3$ and $\lambda $,
and the Higgs mass and top mass are fixed to be 125 GeV and 173.34 GeV, respectively. 
The initial value of these parameters is set to
 be~$y_t(m_t)= 0.937$, $g_1(m_t)=0.358$, $g_2(m_t)=0.648$, $g_3(m_t)=1.167$, and $\lambda(m_t)= 0.126$~\cite{Buttazzo:2013uya} . 

\begin{figure}[ht]
	\centering
	\includegraphics[width=3in]{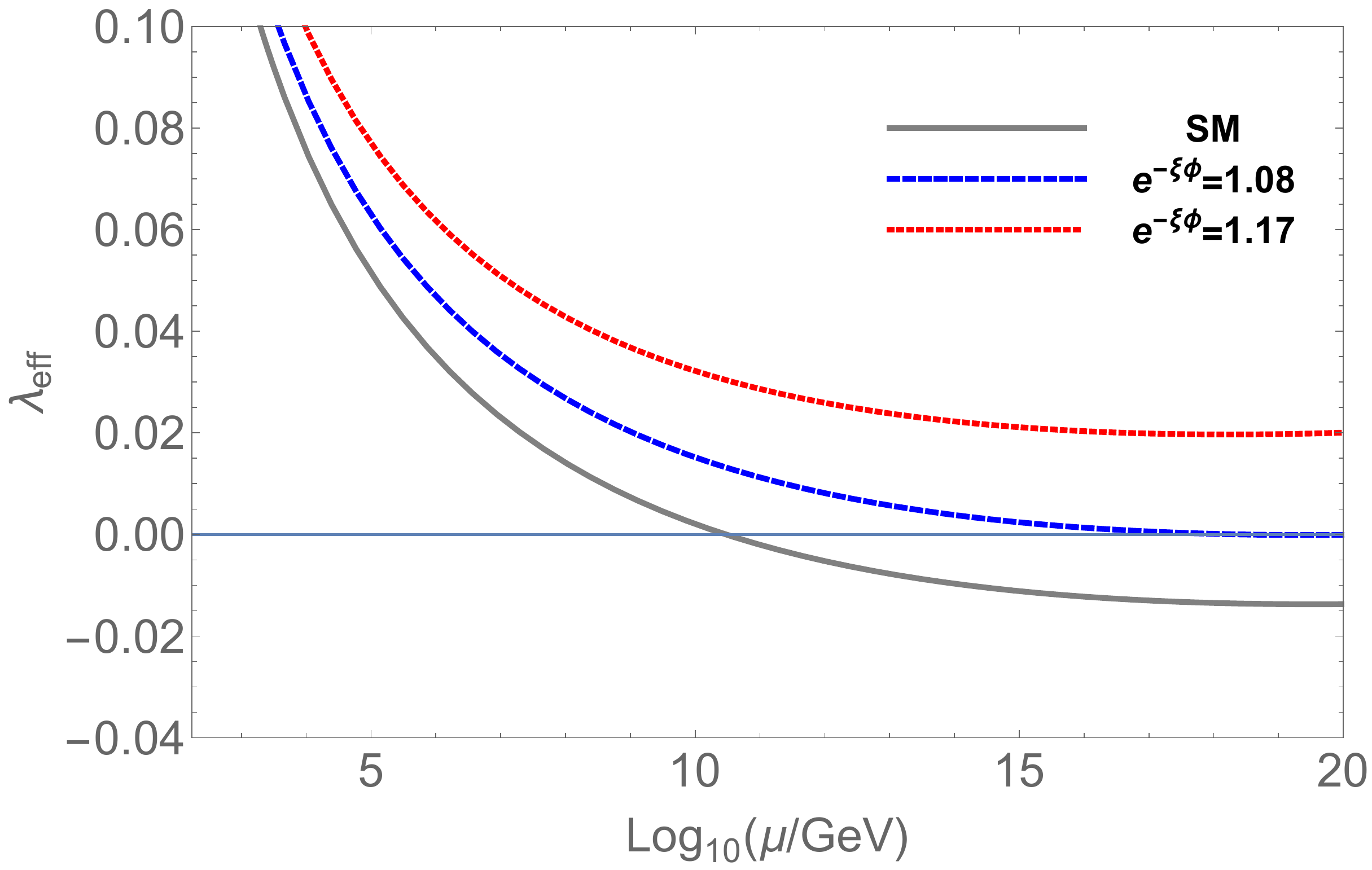}
	\caption{The 2 loop RGE running of $\lambda_\text{eff}$ with different $\xi$ values, where we fixed top mass as the central value 173.34 GeV. 
		The gray solid, blue dashed, and red dotted curves represents $e^{-\xi\phi}=1.0$, $1.08$, and $1.17$, respectively. 
		When $\phi$ during inflation is fixed as is shown in \eqref{main}, the curves correspond to $\xi=0$, $0.35$, and $0.5$,  respectively.}
	\label{plot}
\end{figure}

In Figure \ref{plot} we plot the RGE running of  $\lambda_\text{eff}$ at two loops including the running of the three gauge couplings
 and the top quark Yukawa coupling for different values of $e^{-\xi\phi}$. 
 Note that here we ignore the interactions between $\phi$ and Higgs because it is suppressed by Planck scale. 
 It clearly shows when $e^{-\xi \phi} \gtrsim 1.08$, the electroweak vacuum becomes absolutely stable.
This increase of the self coupling by the quintessence factor $e^{-\xi\phi}$ seems easy to realize in the early universe since $\phi$ runs from a negative value.
 The key point is how large this enhancement could be when we trace back the evolution of quintessence to the era of inflation, 
 and what range of $\xi$ can make the EW vacuum absolutely stable during inflation.

\textit{Evolution of Quintessence}~
Now we study the evolution of the quintessence field $\phi$ in the model \eqref{potential} from the era of inflation until today.
The Friedmann equation and the equation of motion for $\phi$ are
\begin{align}\nn
H^2&=H_0^2\left[\Omega_{\gamma0}\left(\frac{a_0}{a}\right)^4+\Omega_{m0}\left(\frac{a_0}{a}\right)^3\right.\\\label{Friedmann}
&\left.+e^{-\xi\phi}\left(\frac{\lambda\left(|\mathcal{H}|^2-v^2\right)^2}{3\mpl^2H_0^2}+\Omega_{\Lambda0}\right)\right],
\\\label{eom}
0=&\ddot\phi+3H\dot\phi-\xi e^{-\xi\phi}\left(\lambda\left(| \mathcal{H}|^2-v^2\right)^2+\Lambda\right),
\end{align}
where $\Omega_{\gamma0}$, $\Omega_{m0}$, and $\Omega_{\Lambda0}$ are the energy density of the radiation, non-relativistic matter, 
and dark energy at present, normalized by the current critical density $\rho_\text{cr}=3\mpl^2H_0^2$. 
A full analysis of this set of differential equations in the normal quintessence model with potential $e^{-\xi\phi}$ can be found in \cite{Ferreira:1997au,Ferreira:1997hj,Copeland:1997et,Tsujikawa:2013fta}. 
The evolution of this quintessence field may be described by an attractor solution, which is slow-roll in the dark energy dominated universe,
 and ultra-slow-roll in the universe dominated by matter or radiation. We can justify it after solving \eqref{eom} under this assumption.
 To take into account the contribution of the Higgs potential, we approximate it by a step function which jumps from $\Omega_{\Lambda0}$ 
 to $\lambda v^4/\rho_\text{cr}$ at the EW symmetry breaking scale $T_\text{EW}\simeq100\text{GeV}$ as one goes back in time. 
 The value of $\phi$ in the very early universe, i.e., at times well before the EW phase transition, is found to be given by
\begin{align}
e^{\xi\phi}&=\left(\frac{\Omega_{m0}}{\Omega_{\Lambda0}}\right)^{\frac{\xi^2}{3}}
-\frac{\xi^2}{3}\left[1+\mathcal{O}\left(\frac{\lambda}{g_{s*\text{EW}}}
\left(\frac{v}{T_\text{EW}}\right)^4\right)\right],\label{main}
\end{align}
where $g_{s*\text(EW)}\simeq85.5$ is the effective radiation degrees of freedom at the EW scale~\cite{Kolb:1990vq,Husdal:2016haj}. 
The leading order term is from the recent dark energy dominated universe. 
Its main correction from the early universe comes from near the EW phase transition, while all the other contributions are highly 
suppressed like $z_\text{eq}^{-3/2}$ thus completely negligible. 
This $\phi$ should be understood as the initial condition of the quintessence field during inflation. 
During inflation, \eqref{main} contributes to the effective Higgs self coupling, which may prevent it from being negative 
and make it absolutely stable as is discussed earlier. Referring to our result \eqref{main}, we know the EW vacuum will be absolutely stable  
if $e^{-\xi\phi}>1.08\pm 0.02$. Comparing this with the initial value of $e^{\xi\phi}$ given by \eqref{main}, we find that if  $\xi$ satisfies
 \be\label{lowerbound}
 \xi > 0.35 \pm 0.05,
 \ee
 the electroweak vacuum will be absolutely stable before the energy scale reaches $\mpl$, especially during inflation. 
 The uncertainty in \eqref{lowerbound} originates from that of the measurement of the top quark mass, and we have used
  $z_\Lambda\equiv\Omega_{\Lambda0}/\Omega_{m0}\simeq2.21$ from the Planck 2018 data~\cite{Aghanim:2018eyx}. 
  This result is robust in the sense that it is independent of the physics of the early universe, and relies only on the recent 
  expansion history after matter-radiation equality.

\textit{Conclusion}
According to the quantum corrections from all the SM particles, especially from the top quark, the Higgs self coupling becomes negative at around $\Lambda_\text{ins}\sim10^{11}\text{GeV}$. The Higgs instability is a serious problem during the inflationary era in the early universe, 
since the quantum fluctuations of the Higgs boson could have easily exceeded the instability scale. 
We find that in the recently proposed quintessence model where the quintessence field $\phi$ is coupled to Higgs boson 
as $e^{-\xi\phi}(V(\mathcal{H})+\Lambda)$, this problem can be solved. This is because the evolution of $\phi$ can make $e^{-\xi\phi}$ slightly larger
 in the early universe, which can contribute to the effective Higgs self coupling. By requiring that the Higgs EW vacuum is absolutely stable during inflation, 
 we derive a lower bound $\xi\gtrsim0.35$ from the current observational data of dark energy density, matter density, the Higgs mass, the Higgs VEV, 
 and the top quark mass, which is independent of the other parameters and of the physics in the early universe such as phase transitions and inflation. 

The lower bound for $\xi$, together with the upper bound recently obtained by the observational constraints in~\cite{Agrawal:2018own,Heisenberg:2018yae,Akrami:2018ylq,Heisenberg:2018rdu}, $\xi\lesssim0.6$ at 2$\sigma$ CL, sets a testable parameter space 
for the model \eqref{potential}. With Stage-4 surveys such as DESI, LSST, Euclid~\cite{Abell:2009aa,Laureijs:2011gra,Amendola:2012ys,Levi:2013gra}, 
it is possible to constrain $\xi$ up to $\mathcal{O}(0.1)$ within a decade or so.
 This means that whether the model \eqref{potential} can solve the Higgs instability problem could be tested in the near future. 

Recently, the ``Swampland Conjecture'' is proposed~\cite{Brennan:2017rbf,Obied:2018sgi}, which states that all the scalar fields consistent 
with quantum gravity should satisfy $|\nabla V|>cV$, with $c$ being an $\mathcal{O}(1)$ constant. In the current model \eqref{potential}, 
we have $|\nabla V|\simeq\xi V$. Thus the Swampland Conjecture implies $\xi>c$.
For recent discussions on the Swampland Conjecture and its applications to cosmology, see \cite{Agrawal:2018mkd,Agrawal:2018own,Andriot:2018wzk,Dvali:2018fqu,Banerjee:2018qey,Aalsma:2018pll,Achucarro:2018vey,Garg:2018reu,Lehners:2018vgi,Kehagias:2018uem,Dias:2018ngv,Colgain:2018wgk,Brandenberger:2018fdd,Ghalee:2018qeo,Roupec:2018mbn,Andriot:2018ept,Ghosh:2018fbx,Matsui:2018bsy,Ben-Dayan:2018mhe,Chiang:2018jdg,Heisenberg:2018rdu,Damian:2018tlf,Conlon:2018eyr,Kinney:2018nny,Dasgupta:2018rtp,Cicoli:2018kdo,Kachru:2018aqn,Akrami:2018ylq,Nakai:2018hhf,Cho:2018alk,Heisenberg:2018yae,Murayama:2018lie,Marsh:2018kub,Brahma:2018hrd,Choi:2018rze,Quintin:2018loc,Das:2018hqy,Danielsson:2018qpa,Wang:2018duq}.
 Very interestingly, our result \eqref{lowerbound} $\xi\gtrsim0.35$ implies that the Higgs stability and 
 the Swampland Conjecture are satisfied at the same time if $c\lesssim0.35$. 
 Apparently if the Swampland Conjecture is to be satisfied, it should
 be so during inflation as well. The Swampland Conjecture for inflation has been discussed in several recent papers, including the cases of multi-filed inflation~\cite{Achucarro:2018vey}, the curvaton scenario~\cite{Kehagias:2018uem}, $k$-inflation~\cite{Kinney:2018nny}, a non-Bunch-Davies vacua~\cite{Brahma:2018hrd}, and warm inflation~\cite{Das:2018hqy}.  It is therefore not only interesting but also important to see if there exists any consistent model of inflation that smoothly matches to the model considered here. We will leave this issue for future studies.

\textit{Acknowledgment}~
We thank Shinji Mukohyama and Masahito Yamazaki for useful discussions. This work was supported by the 
MEXT/JSPS KAKENHI Nos. 15H05888 and 15K21733, and by the World Premier International Research Center Initiative (WPI Initiative), MEXT, Japan.

\end{document}